\documentclass[
    journal=ancac3,
    manuscript=article
]{achemso}

\usepackage{amsmath,amssymb,amsfonts}
\usepackage{bm}
\usepackage{color, soul}
\usepackage{graphicx}
\usepackage{float}
\usepackage{wrapfig}
\usepackage{verbatim}
\usepackage[english]{babel}
\usepackage[dvipsnames]{xcolor}
\usepackage[
    colorlinks=true,
    bookmarks=false,
    citecolor=blue,
    urlcolor=blue,
    linkcolor=blue,
]{hyperref}
\usepackage{natbib}
\usepackage{subfigure}
\usepackage{amstext, mathrsfs, textcomp}
\usepackage{multirow}
\usepackage{siunitx}
\usepackage{enumerate}
\usepackage{enumitem}
\usepackage{epstopdf}
\usepackage{tabularx}
\usepackage{booktabs}
\makeatletter

\usepackage{physics}
\usepackage{upgreek}

\usepackage{gensymb}

\usepackage{fontawesome}

\DeclareSIUnit\sq{\ensuremath\Box}
\DeclareSIUnit\Kelvin{K}
\sethlcolor{yellow}

\setkeys{acs}{etalmode=truncate,maxauthors=0}
\graphicspath{{Figures/}}

\usepackage{xcolor} % in preamble

\definecolor{mplblue}{HTML}{1f77b4}
\definecolor{mplorange}{HTML}{ff7f0e}
\definecolor{mplgreen}{HTML}{2ca02c}
\definecolor{mplred}{HTML}{d62728}
\definecolor{mplpurple}{HTML}{9467bd}

\usepackage{gensymb}

\DeclareSIUnit\sq{\ensuremath\Box}
\usepackage{xfrac}

\newcommand{\affilEPFL}{Institute of Bioengineering, \'Ecole Polytechnique F\'ed\'erale de Lausanne (EPFL), Lausanne 1015, Switzerland}
\newcommand{\affiITMO}{School of Physics and Engineering, ITMO University, St. Petersburg 197101, Russia}
\newcommand{\affiHEU}{Qingdao Innovation and Development Center, Harbin Engineering University, Qingdao 266000, Shandong, China}
\newcommand{\affiBrescia}{University of Brescia, Brescia, Italy}

\author{Felix Ulrich Brikh}
\affiliation{\affilEPFL}
\author{Aleksei Ezerskii}
\affiliation{\affiITMO}
\alsoaffiliation{\affiHEU}
\author{Olesia Pashina}
\affiliation{\affiITMO}
\altaffiliation{\affiBrescia}
\author{Nikita Glebov}
\affiliation{\affilEPFL}
\author{Wenping Yin}
\affiliation{\affiHEU}
\author{Sergey V. Makarov}
\affiliation{\affiHEU}
\alsoaffiliation{\affiITMO}
\author{Mihail Petrov}
\affiliation{\affiITMO}
\author{Ivan Sinev}
\affiliation{\affilEPFL}
\author{Hatice Altug}
\affiliation{\affilEPFL}
\email{hatice.altug@epfl.ch}

\title{Mid-IR Light Modulators Enabled by Dynamically Tunable Ultra High-Q Silicon Membrane Metasurfaces}

\begin{document}
\begin{abstract}
Metasurfaces have emerged as a powerful platform to control free-space light at the subwavelength scale, enabling applications in sensing, lasing, nonlinear optics, and quantum photonics. However, their practical deployment is hindered by two key limitations: a tradeoff between low-$Q$ resonances and weak amplitude contrast, and their predominantly static nature allowing only passive functionalities. These challenges are further aggravated in the application-relevant mid-infrared (mid-IR) range, where the lack of suitable low-loss materials and the strong absorption of common substrates such as silicon oxide or sapphire severely constrain performance and scalability.
Here, we address these issues with actively tunable single-crystalline silicon membrane metasurfaces that combine high-$Q$ resonances,
strong amplitude contrast, and wafer-scale fabrication compatible dimensions for high throughput manufacturing. Our platform achieves record-high measured $Q$-factors up to 3000 in the mid-IR spectrum, supporting efficient dynamic modulation through two distinct schemes: (i) on-chip electro-thermal tuning \textit{via} Joule heating, sustaining $>\SI{50}{\percent}$ modulation depth at CMOS-compatible voltages and speeds up to \SI{14.5}{\kilo\hertz}, and (ii) ultrafast all-optical modulation \textit{via} carrier generation in silicon, reaching nanosecond response times and estimated sub-GHz modulation rates.
By uniting sharp resonances, strong contrast, large-scale manufacturability, and dynamic tunability, our active silicon membrane metasurfaces advance the frontier of mid-IR nanophotonics and open new opportunities in sensing, free-space communication, thermal radiation management, and quantum technologies.
\end{abstract}

\section{Introduction}
Metasurfaces are ultrathin engineered materials composed of sub-wavelength resonators that can manipulate light with unprecedented precision, enabling control over its phase, spectrum, and polarization~\cite{Yu2014NatMat,Kildishev2013Science,Arbabi2015NatNanotech}. This unique capability allows them to achieve complex optical functionalities efficiently within a compact footprint, matching and surpassing the performance of classical optical devices~\cite{Aieta2012NanoLet,Lin2014Science,Kodigala2017Nature,Lee2017ACSPhot,Corts2022ACSChemRev,Richter2024Jun}. Metasurfaces with ultrahigh quality factor resonances attract particular attention, as their narrowband spectral response and concurrent strong field enhancement enable diverse applications such as efficient lasing\cite{hwang2021ultralow}, high-resolution spectroscopy\cite{tittl2018imaging}, enhanced light–matter interactions\cite{kravtsov2020nonlinear}, and strong nonlinear optical effects\cite{zograf2022high}. High quality factors in metasurfaces have been achieved using the concept of bound states in the continuum (BIC) - perfectly localized optical modes\cite{marinica2008bound}. Although an ideal BIC mode is completely decoupled from the continuum of free space modes, small disturbances, such as symmetry breaking, lead to the introduction of finite radiative losses to the system and the manifestation of sharp resonances in its far-field spectra\cite{sadrieva2017transition}. In particular, quasi-BIC designs that use unit cell symmetry breaking to fine tune the radiative coupling became widely used for precise control of the resonance linewidth\cite{koshelev2018asymmetric}. 

While high-$Q$ resonances strongly enhance performance of static devices~\cite{ouyang2024singular}, many emerging applications require dynamic control over the optical response. Actively reconfigurable metasurfaces~\cite{li2017nonlinear,tonkaev2022multifunctional} allow on-demand modulation of resonance frequency, linewidth, or amplitude, dramatically expanding their functional versatility towards adaptive filters\cite{galarreta_reconfigurable_2020}, LIDARs\cite{berini_optical_2022}, and non-reciprocal optical devices\cite{cotrufo_passive_2024}. Various tuning mechanisms were explored for metasurfaces operating in the visible and near-IR ranges, employing external stimuli such as heat~\cite{Yin2017LightSciApp}, voltage~\cite{Jung2024SciAdv}, photoexcitation~\cite{makarov2015tuning}, or mechanical deformation~\cite{Ou2013NatureNanotech,Tseng2017NanoLett,Saerens2022Nanoscale}. Those approaches rely on changing either the intrinsic material properties of the metasurface (phase transitions~\cite{galarreta_reconfigurable_2020}, carrier concentration~\cite{makarov2015tuning}, piezoelectric~\cite{meng_dynamic_2021}, thermo-optic~\cite{Barulin2024LaserAndPhotRev}, electro-optic~\cite{shirmanesh_electro-optically_2020}, or electro-thermal effects) or the environment (alignment of liquid crystals~\cite{komar_dynamic_2018}, material control through pressure or chemical reactions~\cite{Hail2019AdvOptMat,Gu2022NatPhot}).

\begin{figure}[htb!]
    \centering
    \includegraphics[width=\textwidth]{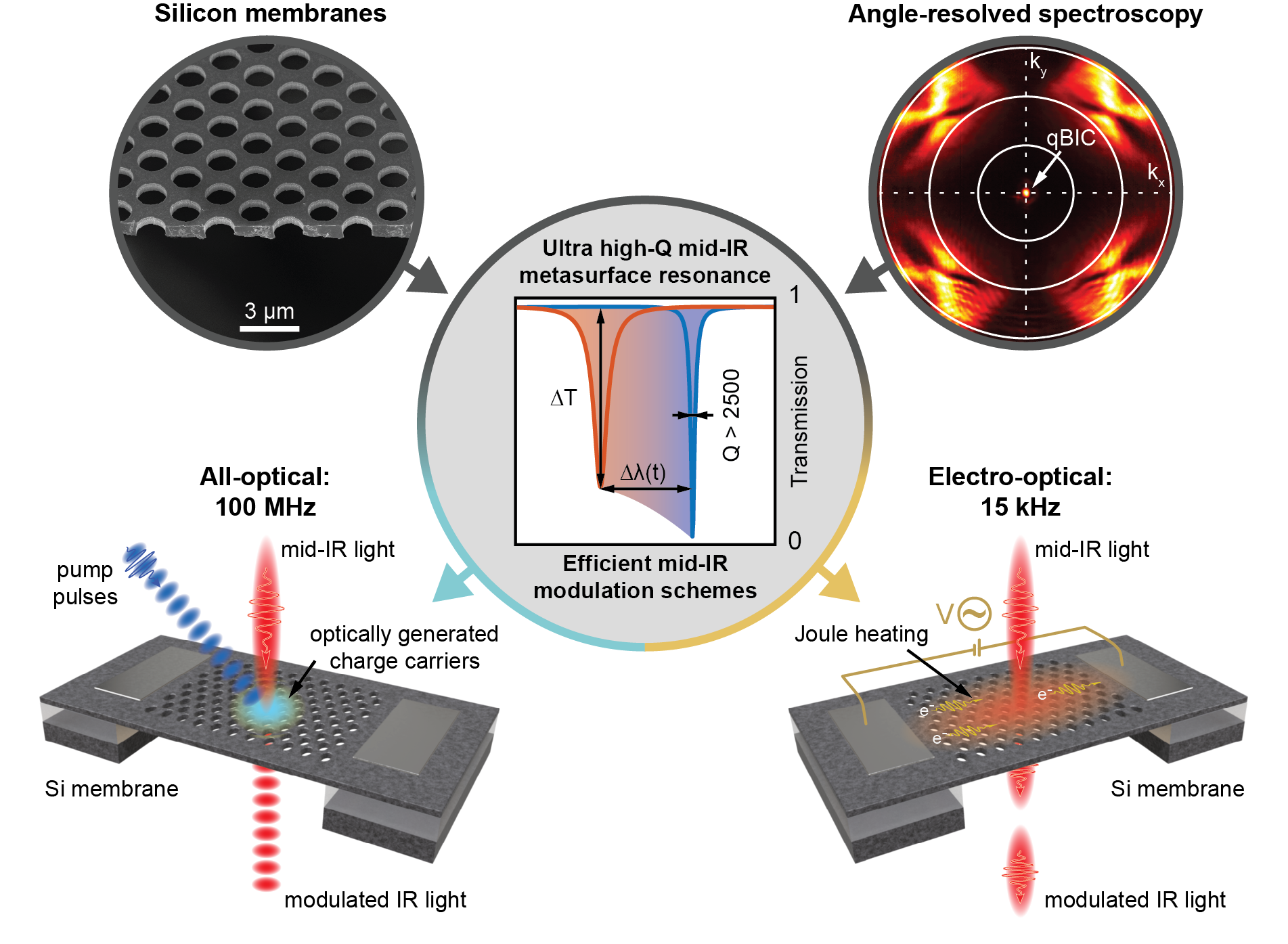}
    \caption{\textbf{Mid-IR metasurface modulators}. Crystalline silicon membrane metasurfaces (top left) in combination with angle-resolved spectroscopy measurements (top right) allow to resolve large amplitude, ultra-high quality factor resonances in mid-IR (middle). This enables efficient mid-IR light modulation schemes. Bottom left: high-speed modulation with optical pump that induces refractive index change through generation of non-equilibrium charge carrier density. Bottom right: electro-optical modulation by direct Joule heating of the membrane at CMOS-compatible applied voltages.}
    \label{fig:Overview} 
\end{figure}

Mid-infrared (mid-IR) photonics have the potential to greatly benefit from both the high $Q$-factor modes and active tuning, which can be utilized for enhanced detection of chemical and biological compounds\cite{tittl2018imaging}, spectroscopy, and efficient high harmonic generation\cite{tonkaev_unconventional_2025}. However, despite the significantly relaxed fabrication requirements for mid-IR metasurfaces
as compared to their visible/near-IR analogues, experimentally reported $Q$-factors are still moderate and do not reach above a few hundreds for neither passive nor active solutions. This is mostly related to the fact that standard fabrication bases, such as silicon-on-insulator and silicon-on-sapphire, that allow to attain $Q$-factors as high as $10^5$ in the telecom spectral range\cite{jin2019topologically}, cannot be scaled towards the mid-IR due to the onset of absorption in the low-index substrate material. At the same time, the commonly used germanium on calcium fluoride platform suffers from grain formation in the sputtered germanium that contributes to the effective non-radiative losses\cite{Richter2024Jun}, limiting the achievable quality factors to $Q\approx200$. Regarding tunability, reconfigurable metasurfaces based on chalcogenide phase-change materials enable strong modulation depth but are limited to low switching speeds, as their tuning mechanism relies on heating and cooling sequences~\cite{Tian2019NatComm,Leitis2020AdvFuncMat,Zograf2021AdvOptAndPhot,S2024ACSPhot}. Similar restrictions arise in devices relying on gas absorption~\cite{Strohfeldt2014NanoLetters}, intercalation of ions~\cite{Ergoktas2021NatPhot} or thermal tuning~\cite{Lewi2018NanoPhot}. On the other hand, electrostatic tuning can potentially provide high (sub-GHz) modulation frequencies in metasurfaces based on graphene\cite{Yao2014NanoLetters,Zeng2018LightSciApp} and ITO~\cite{Park2016NanoLetters}. However, the plasmonic nature of these structures results in even lower $Q$-factors, requiring high operating voltages to achieve reasonable modulation depth\cite{siegel_electrostatic_2024} and rendering them poorly compatible with CMOS standards. This highlights the need for alternative material platforms and design strategies for metasurfaces that can simultaneously support high-$Q$ resonances and efficient, high-speed, low-power tunability in the mid-IR while maintaining compatibility with wafer-scale manufacturing for real-world applications.

In this work, we introduce dynamically tunable high-Q Mid-IR metasurfaces by using single-crystalline silicon membranes as a novel platform. We achieve record-high experimentally measured quality factors with our free-space mid-IR photonic structures. The extremely narrow resonance linewidth combined with large absolute amplitude enable efficient modulation of the metasurface optical response for moderate external drive. We showcase two realizations of actively controlled mid-IR signal modulation (\autoref{fig:Overview}). In the first one, we demonstrate an on-chip device that features strong electro-thermal modulation induced by a CMOS-compatible voltage of \SI{5}{V} sustaining $>50\%$ absolute signal variation up to \SI{14.5}{\kilo\hertz}. In the second, we achieve high-speed (sub-GHz) all-optical modulation using the ultrafast optically induced charge carrier generation in silicon. Our results establish membrane metasurfaces as a versatile platform for active mid-IR light routing that we envision for applications in chemical and biological sensing, thermal emission management, as well as in the emerging fields of mid-IR quantum optics and quantum communications.

\section{Ultra High-$Q$ Resonances in Membrane Metasurfaces}

\begin{figure}[ht]
    \centering
    \includegraphics[width=\textwidth]{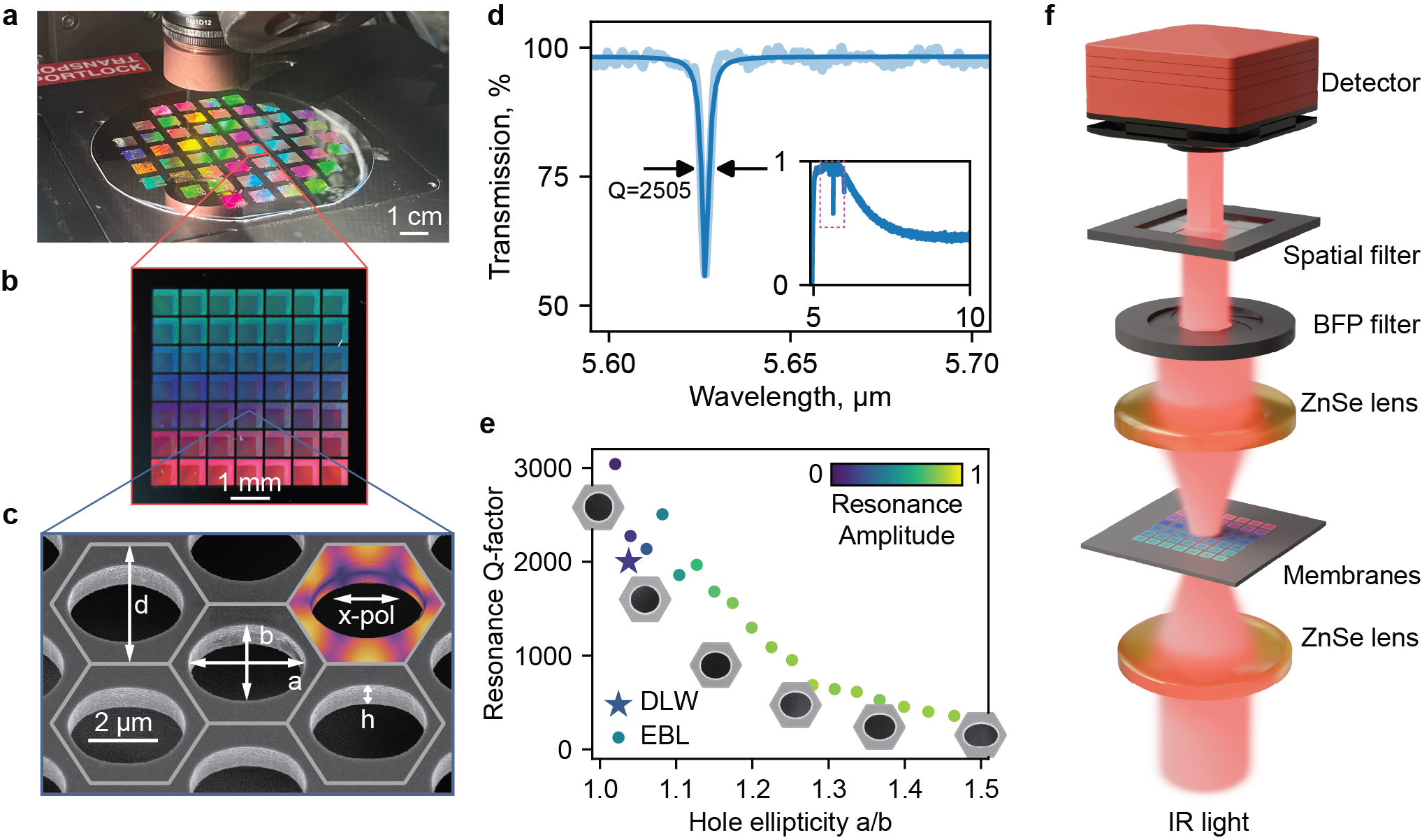}
    \caption{\textbf{Achieving ultra-high quality factors in mid-IR metasurfaces}. (a) Image of wafer-scale membranes fabricated with direct laser writing. (b) Image of a single 1$\times$1 cm$^2$ chip with 7$\times$7 metasurfaces. (c) SEM image of a Si membrane metasurface. Colorful overlay on one of the unit cells shows the calculated distribution of the electric field at the qBIC resonance excited with linearly polarized plane wave at $\lambda_\text{res} = \SI{5.63}{\micro\metre}$. (d) Transmission spectrum of the membrane metasurface with the measured resonance $Q$-factor of $\sim 2505$ and high amplitude contrast. The inset shows the same spectrum in a broader wavelength range. (e) Dependence of the measured $Q$-factor for EBL written membrane metasurfaces with varying hole ellipticity. The insets show SEM images of the corresponding unit cell. The star indicates the performance of DLW written devices. (f) Schematic of the mid-IR spectroscopy setup for high-$Q$ resonance measurements.}
    \label{fig:Metasurface} 
\end{figure}

Our membrane metasurfaces are based on CMOS-grade silicon-on-insulator substrates using high-throughput wafer-scale processes. A metasurface consists of a $h=\SI{1}{\micro\meter}$ thick single-crystalline silicon membrane, perforated by holes arranged in a hexagonal lattice with a period of $d=\SI{3.3}{\micro\meter}$ (for the fabrication details, see Methods). We numerically optimize the structure parameters so that for circular shaped holes, the metasurface supports a symmetry-protected bound state in the continuum in the 5-6~µm spectral range\cite{overvig2020selection,shakirova2025molecular}. This non-radiative state can then be transformed into a mode with a finite $Q$-factor using the quasi-BIC principle\cite{koshelev2018asymmetric}. In our case, we control the radiative losses of the qBIC mode by introducing ellipticity to the holes: $\varepsilon =\frac{a}{b}$ (\autoref{fig:Metasurface}c). To tune the $Q$-factor without a considerable spectral shift of the resonance, when changing the ellipticity we keep the hole area constant at $A = \SI{1.1}{\micro\meter^2}$.

Unlike other designs with small features or multiple resonators per unit cell, our metasurfaces feature large critical dimensions ($\sim \SI{1}{\micro\meter}$) that make them fabrication-friendly, robust to imperfections, and compatible with high-throughput, wafer-scale techniques such as nanoimprint lithography and  direct laser writing (DLW). An image of a 4" wafer fabricated with DLW (see Methods) is shown in \autoref{fig:Metasurface}a.

The crystallinity of the metasurface material and the absence of a substrate that can induce additional optical losses make the silicon membrane-based metasurfaces ideal for achieving ultra-high $Q$-factors in the mid-infrared range. However, direct observation of these high $Q$-factors is also largely dependent on the measurement procedure, as the linewidth and spectral position of the qBIC modes tend to change rapidly with the incidence angle. For accurate measurements of our membrane, we developed a custom FTIR microscope accessory with spatial filtering of signal in the back focal plane, as illustrated in \autoref{fig:Metasurface}f. It allowed us to limit the angular spread of the incident mid-IR light down to several degrees, corresponding to an effective NA of 0.02. \autoref{fig:Metasurface}d shows the transmission spectrum of the metasurface with a hole ellipticity of $\varepsilon=1.08$ fabricated with e-beam lithography (EBL). The spectrum features an extremely sharp resonance dip at around $\lambda$=\SI{5.63}{\micro\meter}. Notably, it combines a very large amplitude contrast close to \SI{50}{\percent} and a high quality factor of $Q=2505$. This represents a major improvement over the previously reported mid-IR metasurfaces, including germanium on CaF$_2$ platform\cite{leitis2019angle} as well as alternative membrane designs\cite{rosas2025enhanced}. By changing the hole ellipticity from $\varepsilon = 1 \text{ to } 1.5$, the $Q$-factors can be tuned in a broad range from several hundreds up to a maximum value of $Q\approx3000$ as shown in \autoref{fig:Metasurface}e. Higher quality factors exhibit smaller modulation depths, which is limited by the residual fabrication imperfections. Structures fabricated with DLW (data marked with star) show comparable $Q$-factors measured at up to 2000, with smaller resonance amplitude due to reduced fabrication fidelity.

\begin{figure}[htbp!]
    \centering
    \includegraphics[width=0.9\textwidth]{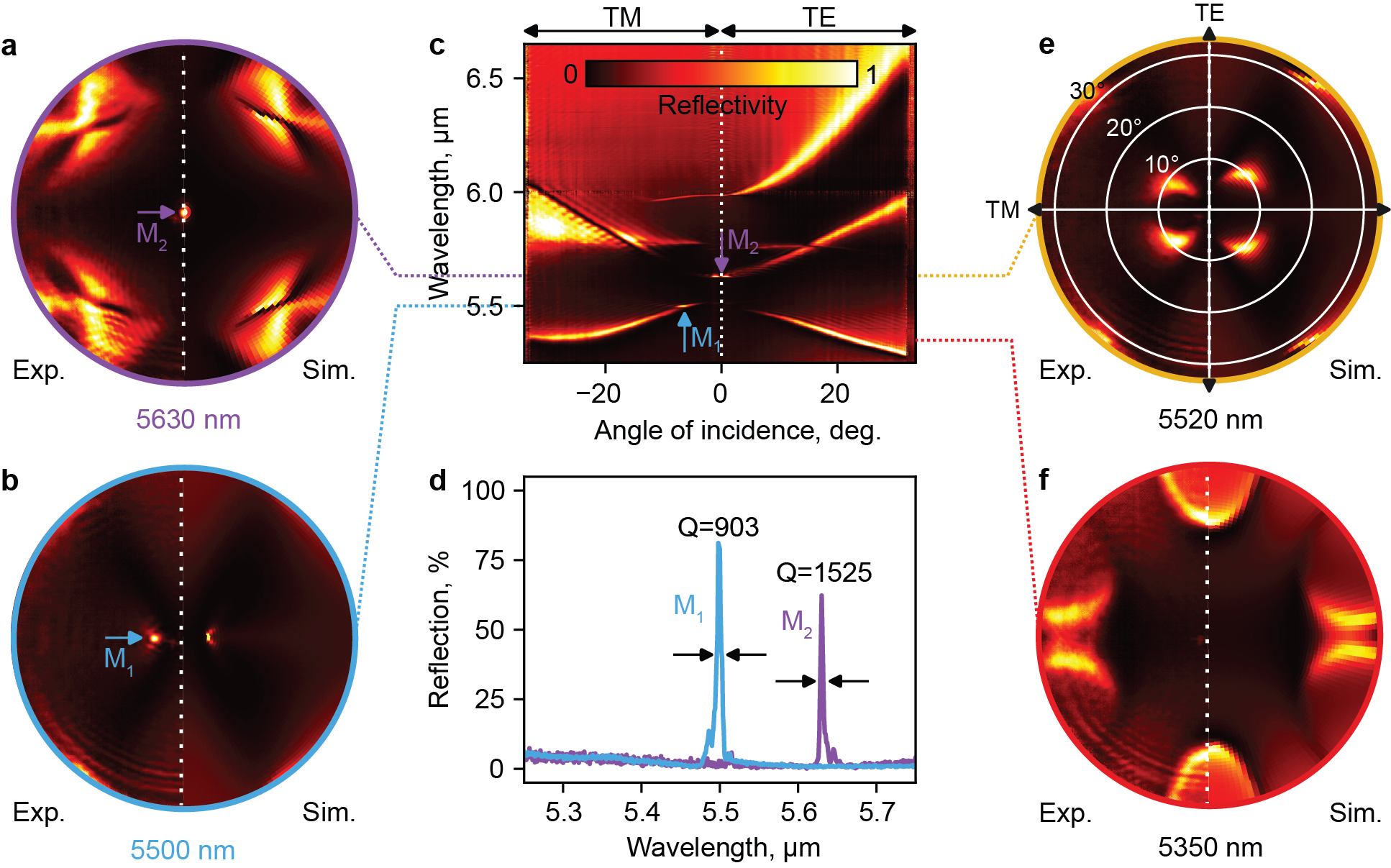}
    \caption{\textbf{Angular and polarization dispersion of membrane metasurfaces}. (a-b) Back focal plane images of the membrane metasurface at two wavelengths (5.63 and \SI{5.5}{\micro\meter}) corresponding to its two high-$Q$ resonances (M$_1$, M$_2$). The left part of each panel shows experimental reflectivity measured with open optics QCL system. The right part shows the simulation data for angle and polarization-resolved reflectivity. (c) Map of angle-resolved reflectivity for TE and TM polarization extracted from the experimentally measured back focal plane images. High-$Q$ modes M$_1$, M$_2$ are indicated with arrows. (d) Metasurface reflectivity resonances extracted from the measured back focal plane data. The $Q$-factors are modified with respect to the extracted valued from customized FTIR measurements due to the finite imaging quality of the BFP setup. (e,f) Back focal plane images of the membrane metasurface for off-resonant wavelengths (5.52 and \SI{5.35}{\micro\meter}). The overlay in panel (e) shows the polar axes common for images in a,b,e,f}
    \label{fig:BFP} 
\end{figure}

To further characterize the optical properties of high-$Q$ membrane metasurfaces, we developed a mid-IR back focal plane (BFP) open optics imaging setup (see Methods section for more details). It is based on a quantum cascade laser assembly (from Daylight Solution Spero) that provides a narrow ($<0.01$~cm$^{-1}$) emission line tunable in a broad spectral range~(5-10~µm). In the setup, we focus the laser radiation on the sample surface through a 0.56~NA black diamond lens, accommodating incident angles up to $\sim 34\degree$. We then record the reflectivity images of the back focal plane of the system, projected on a mid-IR microbolometer camera (DataRay IR-BB) using a $4f$ lens assembly. Collecting these images while tuning the laser wavelength, we acquire polarization- and angle-resolved hyperspectral reflectivity data that manifests the isofrequency curves of the metasurface modes. Four exemplary back focal plane images (normalized to reflection images from a gold mirror) are presented in \autoref{fig:BFP}a,b,e,f. The right half of each plot is replaced with the corresponding BFP data obtained with numerical simulations and shows perfect agreement with the experiments. In particular, \autoref{fig:BFP}a and b show the BFP images at wavelengths close to the high-Q resonances of the metasurface, which feature sharp peaks with extremely narrow angular spread.

We further process the hyperspectral BFP data to extract the dispersion of the optical modes of our metasurface. \autoref{fig:BFP}c shows the reflectivity maps for pure TE and TM polarization of the incident light, obtained by stacking the vertical and horizontal middle sections of each recorded BFP image. Two dispersive high-$Q$ modes (marked M$_1$ and M$_2$) are clearly visible, with M$_2$ corresponding to the qBIC mode that strongly manifests for TE~polarization. Vertical sections in \autoref{fig:BFP}c represent the reflectivity spectra at selected incidence angles, while \autoref{fig:BFP}d highlights the spectra of modes M$_1$ and M$_2$. From this data, we extracted  $Q$-factors of $Q_1=903$ and $Q_2=1525$. The smaller $Q$-factors, compared to measurements with the customized FTIR, can be attributed to the aberrations and limited imaging performance of our BFP set-up. These effects reduce the measured $Q$-factors by broadening the detected peaks, analogous to averaging over a finite angular range.

\begin{figure}[htbp!]
    \centering
    \includegraphics[width=\textwidth]{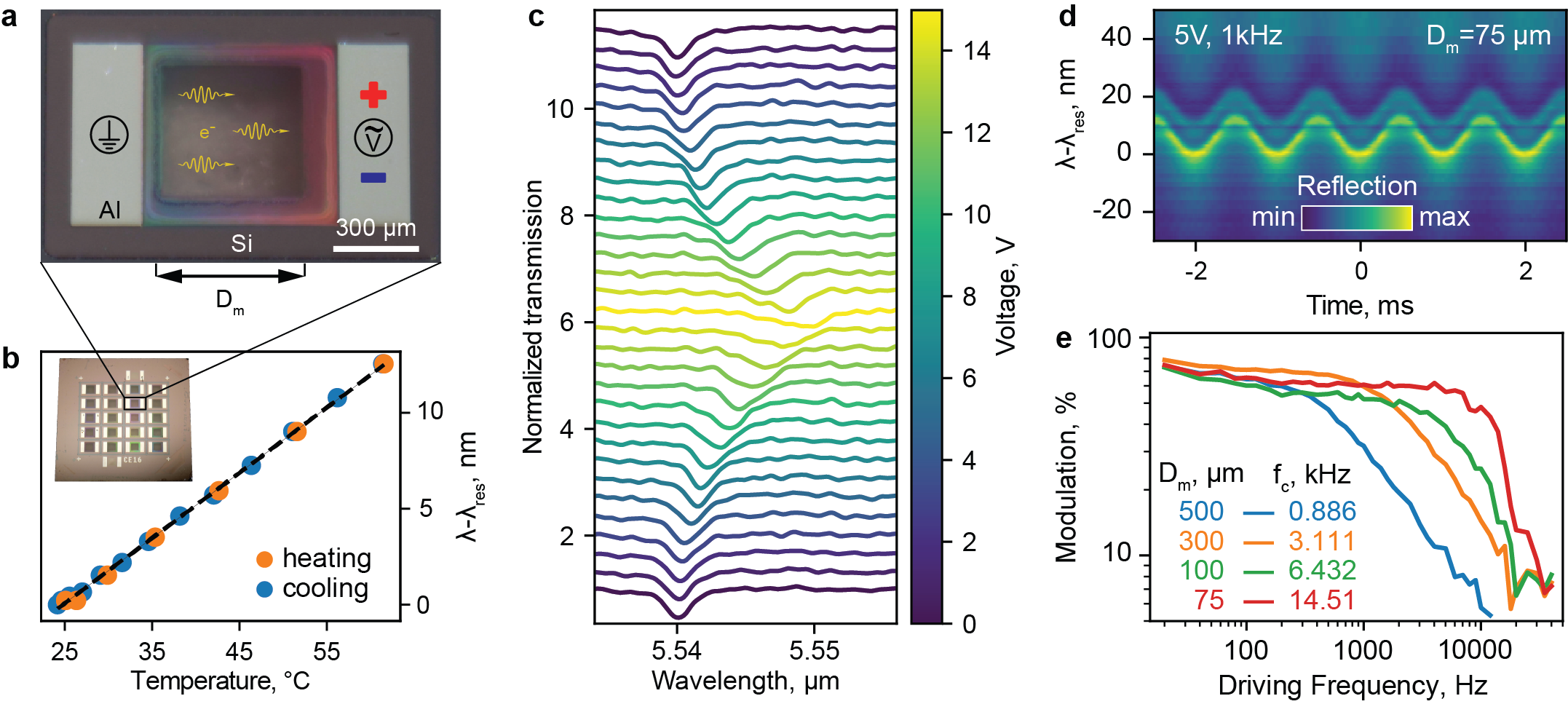}
    \caption{\textbf{Electro-thermal modulation}. (a) Image of the mid-IR electro-optical modulator, consisting of a membrane metasurface with a size D$_\text{m}$ and two aluminum electrodes. The overlay visualizes the Joule heating induced by the current passing through the membrane. (b) Dependence of the measured resonance wavelength shift on the membrane temperature. The inset shows an image of a $9\times9$~mm$^2$ chip featuring 20 electro optical devices. (c) Dependence of a \SI{500}{\micro\meter} metasurface transmission spectra on the applied static voltage (encoded by the color of each curve). (d) Modulation of the spectral position of the metasurface reflectivity resonance recorded in time domain with the back focal plane imaging setup. (e) Amplitude-frequency characteristics for metasurface modulators with different membrane sizes $D_m$ measured at \SI{5}{\volt} driving voltage. The inset shows the cutoff frequency $f_c$ for each case.}
    \label{fig:ElMod} 
\end{figure}

\section{Electro-Thermal Modulation}

After establishing Si membrane metasurfaces as a platform supporting ultra high-$Q$ resonances in the mid-IR, we leverage them to achieve active mid-IR light modulation with high contrast in two complementary ways. For this, we rely on the outstanding properties of silicon: firstly, its high thermo-optical coefficient for on-chip electro-thermal tuning and secondly, the strong dependence of its refractive index on the charge carrier density for fast all-optical tuning. 

To realize electro-thermal modulation, we implement direct Joule heating of the membrane by passing electrical current through it. For this purpose, we evaporate thin-film aluminum electrodes onto the opposite sides of the metasurface, thus enabling the current flow with minimal perturbation of the optical response (see \autoref{fig:ElMod}a). To facilitate sufficient heating efficiency at CMOS-compatible voltages, we lightly dope the silicon membrane with phosphorous (see Methods for details). This introduces a trade-off: while doping reduces the electrical resistance and enables more efficient heating at low voltages, it also increases the non-radiative optical losses, eventually deteriorating the resonance quality factor. The doping level was therefore carefully optimized to balance the electrical conductivity with optical performance.

High quality factor and large resonances amplitude of our membrane metasurfaces lead to efficient tuning, as the relatively high thermo-optical coefficient of silicon still measures at only \SI{2e-4}{\per\Kelvin}. Therefore, the extremely narrow resonance linewidth facilitates a strong modulation depth for minimal temperature changes. This not only increases the tuning efficiency, but is also critical for reaching higher modulation speeds, for which the heat dissipation becomes the main challenge.

To calibrate the electro-optical tuning, we first performed the spectroscopy experiments with external heating of the metasurface using a ceramic hotplate in a temperature range of 25 to 60$^{\circ}$C (\autoref{fig:ElMod}b). We then applied a series of static voltages across the metasurface and recorded the corresponding transmission spectra. The results reveal a clear shift of the resonance with the increase of the applied voltage (\autoref{fig:ElMod}c). The shift scales linearly with the temperature (\autoref{fig:ElMod}b) and quadratically with the applied voltage, consistent with the expected linear relationship between the resonance wavelength and the heating power. Remarkably, a temperature change of only 35$^{\circ}$C leads to a resonance shift of up to 5~FWHM ($\approx\SI{11}{\nano\meter}$ for $Q=2500$  and $\lambda_\text{res}=5.63$~µm) and an absolute modulation amplitude of 70$\%$.

We then studied the dynamic performance of the electro-optical tuning mechanism by modulating the applied voltage. The long acquisition time of FTIR spectra prevents the observation of fast resonance tuning. Therefore, we used our open optics setup with QCL excitation (see Methods for details) and applied a spatial filter in the Fourier plane of the optical system to isolate the response of the high-$Q$ resonance mode. In the experiment, the voltage-induced refractive index change steers the metasurface reflectivity maximum away from the aperture of the spatial filter. The corresponding signal change is then measured with a fast mercury cadmium telluride (MCT) detector. By scanning the excitation wavelength and tracking the detector readout with an oscilloscope, we can track the temporal modulation of the metasurface resonance \autoref{fig:ElMod}d. With a driving voltage of \SI{5}{\volt} at 500~Hz, the resonance clearly follows the sinusoidal curve with a frequency of \SI{1}{\kilo\hertz}, as expected from the quadratic dependence of dissipated power on voltage.

As mentioned above, the main limiting factor for electrically driven suspended membrane systems is the slow heat dissipation because of the poor thermal conductivity of air. The restricted thermal flow hinders the cooling rate and, consequently, the modulation speed. We address this issue by fabricating smaller membranes, which have more efficient lateral heat dissipation. However, when reducing the membrane size one needs to keep in mind that the BIC mode used in our design is of a collective nature, and requires a minimum number of unit cells to maintain its ultra-high $Q$-factor and resonance amplitude. Earlier studies on this subject reported the lateral spread of qBIC modes within up to 20 unit cells~\cite{Liu2019PRL,Dong2022LightSciApp,Goelz2024AdvMat}. In our case, we achieve 10-fold higher $Q$-factors, indicative of even larger mode volumes. Based on this analysis, we fabricated metasurface membranes with lateral sizes of $D_\text{m} =$~\SIlist{500;300;100;75}{\micro\meter}, and studied their amplitude-frequency response. \autoref{fig:ElMod}e shows that all of the metasurfaces exhibit a base modulation depth of \SI{70}{\percent}, confirming that their size is sufficient to support an unperturbed high-$Q$ optical mode. At the same time, the reduction of the membrane size drastically improves the cut-off frequency due to more efficient lateral heat dissipation that allows faster membrane recovery. The maximum electro-optical modulation speed reached \SI{14.5}{\kilo\hertz} for $D_\text{m} = \SI{75}{\micro\meter}$ membrane, with the total dissipated power of \SI{32}{\milli\watt}.

\begin{figure}[htbp!]
    \centering
    \includegraphics[width=\textwidth]{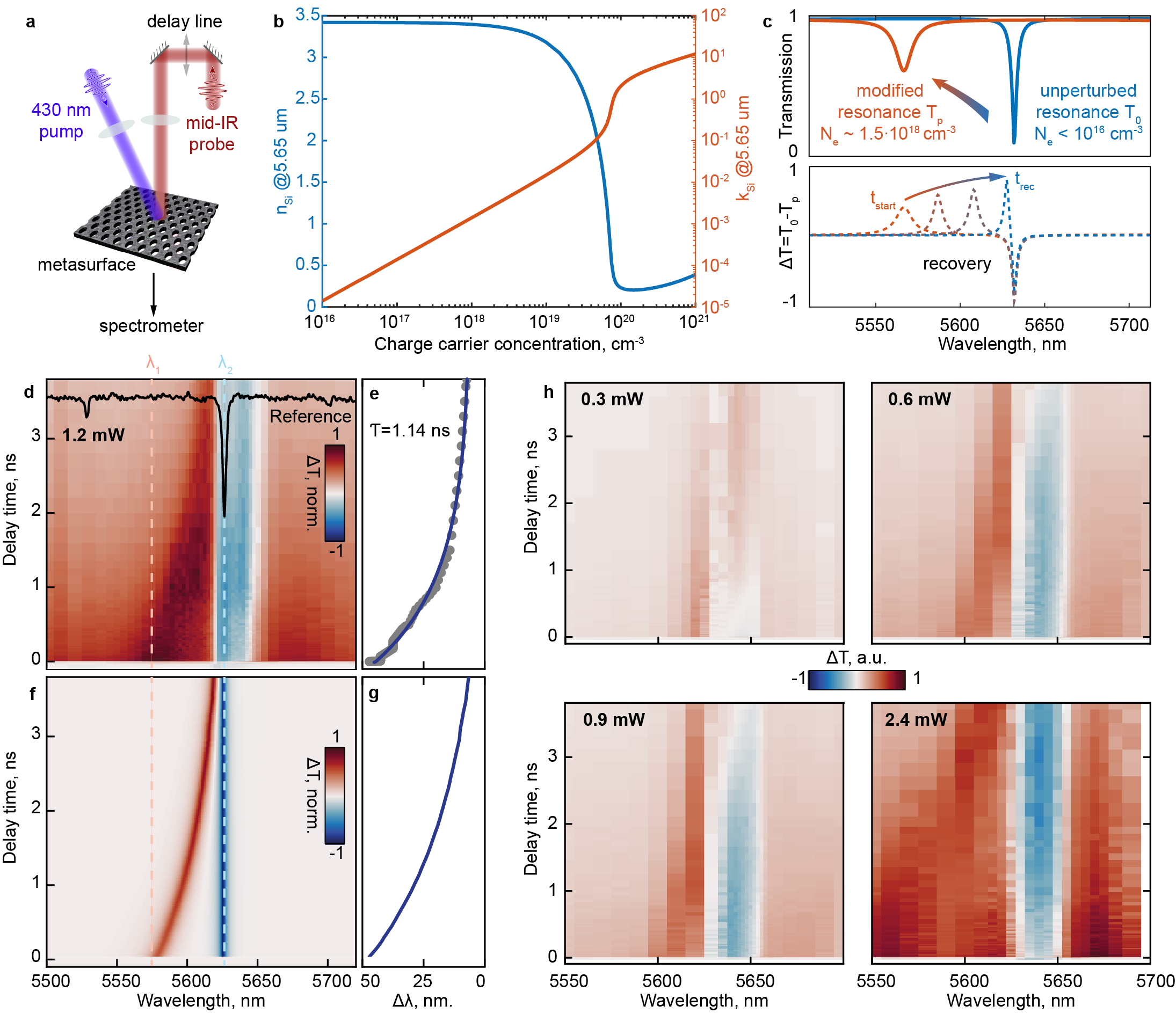}
    \caption{\textbf{Ultrafast all-optical modulation}. (a) Schematic of the pump-probe experiment for studying of the temporal dynamics of membrane metasurfaces. (b) Calculated dependence of the real and imaginary part of the refractive index of crystalline silicon at $\lambda$=\SI{5.65}{\micro\meter} on the charge carrier density $N_e$. (c) Illustration of the all-optical tuning concept. Top: the narrow transmission resonance of the unperturbed metasurface (blue curve) is blue-shifted and broadened (red curve) when the incoming pump pulse induces the non-equilibrium charge carrier density. In the spectrum of differential transmission $\Delta$T (bottom), this leads to the formation of a characteristic asymmetric profile with a dip at the original resonance position, which gradually recovers over time. (d) Temporal dynamics of the normalized differential transmission of the membrane metasurface for pump-probe delay time range of 0-3.6~ns. The metasurface is pumped optically at $\tau$=0 with 430~nm, 290~fs pulse at 10~kHz repetition rate with an average power of \SI{1.2}{\milli\watt}. The black curve in the overlay shows the transmission spectrum of the unperturbed metasurfaces measured with FTIR. (e) Temporal evolution of the resonance shift. The dots show data points extracted from experiment, curve is the fitting of data with a single exponent with $\tau$=1.14~ns. (f,g) Time-resolved dynamics of the metasurface resonance calculated assuming a decay rate of $N_e$ of 2.28~ns. (h) Measured time-resolved differential transmission of the metasurface under different pump powers indicated in each panel.}
    \label{fig:OptMod} 
\end{figure}

\section{All-Optical Modulation}

To achieve fast modulation of Mid-IR light, we utilized optically induced charge carrier generation. This mechanism was previously successfully used for tuning the optical response of silicon photonic structures in the visible~\cite{makarov2015tuning,sinev2021observation}. Mid-IR range is particularly interesting for this all-optical tuning approach, as the increase of the charge carrier density blue-shifts the silicon plasma frequency close to the spectral range of the resonance. This leads to a strong modulation of the refractive index at moderate charge densities, minimizing the power consumption. Here, we achieve strong modulation for laser fluences less than $\SI{5}{\micro\joule\per\centi\meter\squared}$, almost 2 orders of magnitude lower than what is required for visible and near-IR applications\cite{sinev2021observation,aigner2025optical}

In our experiments, we employed visible light ($\lambda=\SI{430}{\nano\meter}$) pulses to generate non-equilibrium charge carriers in the silicon membrane, thus altering its refractive index in an ultrafast manner. We then probed the modifications to the metasurface response using a spectrally tunable mid-infrared probe pulse with a controlled time delay with the probe (\autoref{fig:OptMod}a). We estimated the expected refractive index change with simulations of the charge-density-dependent properties of silicon, taking into account the plasma frequency shift, band-filling effect, and changes to the band gap. The simulation results for the real and imaginary parts of the refractive index at the resonant frequency of the metasurface ($\lambda$=\SI{5.65}{\micro\meter}) are shown in \autoref{fig:OptMod}b. They reveal that the real part $n$ exhibits a gradual decrease with the increase of $N_e$, accompanied by a steady growth of loss $k$. At a moderate charge density of $N_e=\SI{7.4e19}{\per\centi\meter\cubed}$, silicon is metalized in this spectral range as its permittivity drops below zero. 

The concept of ultra-fast all-optical modulation of the high-$Q$ membrane is shown in \autoref{fig:OptMod}c. The spectrum of the unperturbed metasurface ($T_0$) exhibits a sharp transmission dip (blue curve). With the arrival of the pump pulse, the generated charge carriers decrease the refractive index of the membrane material (\autoref{fig:OptMod}b), which leads to a spectral blueshift of the resonance ($T_p$). In the differential transmission spectrum $\Delta T = T_0-T_p$ (dashed curves), this manifests itself as a bisignate profile, which then merges together during the subsequent charge recombination process (decay of $N_e$).

The charge recombination time is a critical parameter that defines the membrane recovery time. Bulk crystalline silicon has low defect density, which extends the charge lifetime. Our simulations indicate that the surface recombination process dominates the charge decay at the estimated peak concentration values, while the Auger process becomes relevant for concentrations of at least an order of magnitude larger. Furthermore, in the experiment, patterning the membrane significantly increases the surface area, further facilitating the corresponding charge recombination mechanism. We measure the decay time directly in pump-probe experiments by using a long (4~ns) delay line and tracking the recovery of the metasurface resonance in differential transmission spectra. \autoref{fig:OptMod}d shows the map of $\Delta T$ measured for the average pump power of 1.2~mW (equivalent laser fluence $\SI{4.8}{\micro\joule\per\centi\meter\squared}$). As expected, the arrival of the pump pulse at (probe delay $\tau$=0) induces a characteristic bisignate pattern in the spectrum, with a dip at the wavelength of the unperturbed resonance and a blueshifted peak corresponding to the optically modified mode (see also illustration in \autoref{fig:OptMod}c). Note that the peaks in the probe pulse transmission spectra are broadened as compared to linear spectroscopy data, which is shown with a black curve in \autoref{fig:OptMod}d. This is due to the divergence of the focussed probe beam, leading to the averaging of the response over multiple angles. Increasing $\tau$ reveals a gradual recovery of the transmission profile governed by the decay of $N_e$. The metasurface resonance shifts from the maximally detuned position $\lambda_1$ to the unperturbed value $\lambda_0$, both marked with vertical dashed lines in the map. From these data, we extracted the dependence of the resonance peak shift on the probe delay time, which is shown in \autoref{fig:OptMod}e. Fitting the data with a single exponent yields a maximum detuning of 50~nm and the characteristic decay time of the resonance shift of 1.14~ns. As the resonance shift has almost linear dependence on the refractive index, and, consequently, a square root dependence on the concentration $N_e$, we estimate the carrier concentration decay time at \SI{2.28}{\nano\second}. We then performed numerical simulations of the time-resolved metasurface transmission spectra for a broad range of concentration values using the data from \autoref{fig:OptMod}b. For each time delay value, we retrieved the corresponding charge concentration by fitting the calculated resonance wavelength to the transient transmission maxima extracted from the experiment (\autoref{fig:OptMod}e). The fitting yielded a maximum concentration of $N_e=\SI{1.5e18}{\per\centi\meter\cubed}$ and a lifetime of \SI{2.25}{\nano\second} consistent with the fitting of the resonant shift lifetime (\autoref{fig:OptMod}f,g). According to our estimations, this is fully determined by a rather high surface recombination velocity ($\SI{5e4}{\centi\meter\per\second}$), assisted by the membrane patterning, while Auger processes are expected to manifest for at least an order of magnitude higher initial concentrations. The transient transmission spectra measured for different pump powers (\autoref{fig:OptMod}h) indicate that lower pump powers naturally lead to smaller initial resonance shifts, with bisignate lineshape almost vanishing for the pump power of 0.6~mW. For the maximum available pump power of 2.4~mW, we observed resonance detunings of up to 80~nm, which corresponds to the induced carrier concentrations of $\approx \SI{5e18}{\per\centi\meter\cubed}$.

The short decay time of the carrier concentration facilitates high modulation speeds. To estimate their limits, we consider both the temperature stability of the membrane and the charge relaxation. The data on electro-optical tuning indicated that the metasurface could sustain dissipation of at least 32~mW of power. For the membrane size of \SI{75}{\micro\meter}, which is able to support the high-$Q$ resonance, this dissipated power is equivalent to a complete absorption of the $\SI{4.8}{\micro\joule\per\centi\meter\squared}$ laser pump pulse at a repetition rate of $\approx$120~MHz. On the other hand, at this modulation frequency, the resonance shift $\Delta\lambda$ decays down to 2~nm before the arrival of the next pulse. This puts the detuning below the resonance FWHM, indicating sufficient recovery of the transmission amplitude. %\commentVanya{Will add a comparative analysis with other modulation solutions.}

\section{Conclusion}
We have demonstrated the vast potential of single-crystalline silicon membrane metasurfaces as a versatile platform for active mid-IR photonics. The crystalline material quality, combined with the absence of a lossy substrate, enables record-high measured resonance quality factors up to 3000, over an order of magnitude higher than previously reported metasurfaces based on amorphous materials. The narrow resonance linewidth, together with strong amplitude contrast and fabrication-friendly critical dimensions, provide the foundation for dynamic and scalable mid-IR photonic devices.

Building on this platform, we implemented two complementary modulation schemes. Electro-thermal tuning achieves large modulation depths exceeding 50\% at CMOS-compatible voltages, while scaling to modulation speeds of up to \SI{14.5}{\kilo\hertz} through  membrane size optimization. In parallel, all-optical modulation \textit{via} laser-induced charge carrier generation provides nanosecond response times, supporting estimated sub-GHz operation at low pump fluences. These approaches overcome the trade-off between large modulation amplitude and high modulation speeds.

By uniting high resonance $Q$-factors with strong amplitude contrast, efficient dual-mode tunability and wafer-scale fabrication, our silicon membrane metasurfaces address the important bottlenecks in mid-IR photonics technologies. This paves the way for CMOS-compatible, active devices for applications in chemical and biological sensing, free-space communications, thermal emission management, and quantum optics. Furthermore, our developed platform opens new avenues for future advances in dynamically reconfigurable photonic systems.

\section{Methods}\label{sec:Methods}
\subsection{Device Fabrication}
Silicon (\text{Si}) membrane metasurfaces were fabricated starting from silicon on insulator (SOI) wafers: \SI{1}{\micro\meter} device Si (boron doped \SIrange{8.5}{11.5}{\ohm\cdot\centi\meter}) - \SI{1}{\micro\meter} silicon dioxide (SiO$_2$) buried oxide - \SI{250}{\micro\meter} handle Si. First, the backside openings were defined lithographically in a \SI{3}{\micro\meter} SiO$_2$ layer (Plasma-Therm Corial D250L PECVD) using direct laser writing (DLR) (Heidelberg Instruments Maskless Aligner MLA 150, AZ ECI 3027, 3000~rpm) and fluorine based deep reactive ion etching (DRIE, SPTS Advanced Plasma System). Second, the device Si layer is patterned with a single-step lithography process (DLR: Heidelberg Instruments VPG200, AZ ECI3007, 6000~rpm; \textbf{or} EBL: Raith EBPG5000, ZEP520 50\%, 2000~rpm) followed by DRIE (Alcatel AMS 200 SE). Third, the membranes are opened by etching the handle silicon through the previously defined SiO$_2$ mask using a DRIE Bosch process (Alcatel AMS 200 SE). Lastly, the buried oxide layer is removed from the opening areas by Hydrofluoric acid (HF) vapor etching (SPTS µEtch).

\subsubsection{Doping}
For electro-optical modulation experiments, the device silicon was doped with phosphorus in vapor phase (Centrotherm furnace, \SIrange{1}{4}{\minute} POCl$_3$ at \SI{800}{\celsius}, HF deglazing, \SI{30}{\minute} annealing at \SI{900}{\celsius}). The achieved sheet resistance range of \SIrange[per-mode=symbol]{44.8}{347}{\ohm\per\sq} was enough to find a suitable trade-off between optical and electrical properties. The contact electrodes were defined in photoresist (AZ ECI 3027, 3000~rpm) by direct laser writing ($\lambda = \SI{405}{\nano\meter}$, \SI{200}{\milli\joule\per\centi\meter\squared}). After development (AZ 726 MIF, 2~min) and descumming in O$_2$ plasma (\SI{500}{\watt}, \SI{1}{\minute}), the native SiO$_2$ was stripped by 10~s dip in 7:1 buffered HF followed by electron beam evaporation (Alliance-Concept EVA 760) of 20~nm titanium and 150~nm aluminum. The final step was the lift-off of the superfluous metal (MICROPOSIT Remover 1165). 

\subsection{Infrared spectroscopy}
\subsubsection{FTIR Microscope}
We obtained the infrared (IR) transmission spectra at normal incidence using a Bruker Vertex 80v FT-spectrometer with an attached IR Microscope (HYPERION 3000) equipped with a liquid nitrogen cooled MCT detector. The metasurfaces were excited using a ZnSe lens with the focal length of 50~mm mildly focusing linearly polarized IR light on the sample surface. Transmitted light was collected with another 25~mm lens equipped with an additional iris placed at its back focal plane. Closing the iris allowed for limiting the numerical aperture of the system down to approximately 0.02 and thus suppressing the unwanted signal from oblique excitation angles. Signal collection area was limited to a approximately \SI{300}{\micro\meter} square central region of the membrane by a double-blade aperture placed in the conjugate image plane of the IR microscope. The sample chamber was constantly purged with dry air to provide stable low level of humidity. 

\subsubsection{Free Space BFP Setup}
The light source for the back focal plane measurement is a tunable QCL laser (Daylight Solutions SPERO QTX 340). The light was focused onto the metasurface using a black diamond lens (f=\SI{5.95}{\milli\meter}, NA=0.56). The reflected light was collected and spatially filtered in the real and Fourier planes of a 4$f$ system. The signal was then either imaged on a mircobolometer based camera with 600$\times$480 pixels (DataRay WinCamD-IR-BB) or detected with a MCT photodiode (Thorlabs, PDAVJ10).

\subsubsection{Pump-probe measurements}
Ultrafast dynamic measurements were conducted using a pump-probe system (Harpia-TA, Light Conversion) with a \SI{4}{\nano\second} delay line. The femtosecond laser pulses (\SI{1030}{\nano\meter}, \SI{200}{\femto\second}, \SI{50}{\kilo\hertz}), generated by the Pharos PH2-20W laser system (Light Conversion), were split for the pump and probe beams. Each beam was individually tuned using an optical parametric amplifier (Orpheus-HP for the pump and Orpheus-One-HP for the probe). 
Both beams were focused on the sample, which was positioned orthogonally to the probe beam.
%The overlap of the pump and probe beams was ensured initially by visual positioning using an IR detection card and then fine-tuned to optimize the signal at the peak corresponding to the detuned resonance position ($\lambda_1$ in \autoref{fig:OptMod}d) at a \SI{100}{\pico\second} delay.
The infrared transmission spectra were collected using an Andor KYMERA-193i-B2 spectrograph.
For the average (chopped) measured pump power of \SI{1.2}{\milli\watt} and an approximate beam area of \SI{1}{\milli\meter\squared}, we estimate a fluence of \SI{4.8}{\micro\joule\per\centi\meter\squared}. 

\subsection{Data Processing}
\paragraph{FTIR}
Spectral data from the FTIRs, normalized to signal without sample with the same optics, was fitted with a Fano lineshape to obtain resonance frequency, $Q$-factor and amplitude.

\paragraph{IR camera}
Back focal plane reflectivity data from the 2D bolometric detector was normalized to the reflection BFP image of a gold mirror for each excitation wavelength.

\paragraph{Photodiode}
The tunable QCL provides pulses of \SIrange{20}{500}{\nano\second} length at a repetition rate of up to \SI{100}{\kilo\hertz}, resulting in a maximum duty cycle of \SI{5}{\percent}. The pulse train was collected with an oscilloscope (Agilent DSO-X 3012A). The oscilloscope sampling rate was set between \SIrange{10}{400}{\mega Sa \per\sec}, which allowed to resolve each pulse with a minimum of \num{5} samples while simultaneously recording down to \SI{10}{\hertz} modulation.
The pulse train was converted to intensity (\autoref{fig:ElMod}d) by averaging the oscilloscope readout with a rolling window matching the pulse length, followed by max pooling with a window matching the laser repetition rate.

\subsection{Simulations}
The metasurface spectra were simulated using the Frequency Domain Solver in CST Studio Suite with plane wave excitation. To calculate the back focal plane images, we did sweeps of the inclination and azimuthal angles of the incident k-vector with account for the symmetry of the structure.

Simulations of electromagnetic field distribution, thermal effects, and carrier dynamics were performed using COMSOL Multiphysics.

\section*{Acknowledgements}
We acknowledge the European Union's Horizon Europe Research and Innovation Programme under agreements 101046424 (TwistedNano) and 101070700 (MIRAQLS). This work was supported by the Swiss State Secretariat for Education, Research and Innovation (SERI) under contract numbers 22.00018 and 22.00081. F.B., N.G. and I.S. thank Gloria Davidova for support with wafer-scale fabrication. The authors acknowledge the use of nanofabrication facilities at the Center of MicroNano Technology of \'Ecole Polytechnique F\'ed\'erale de Lausanne. O.P. and M.P acknowledge the the support of the work by the Federal Academic Leadership Program Priority 2030. This research was also supported financially by the Overseas Outstanding Youth of Shandong Province (Grants 2024HWYQ-082). S.M. acknowledges the support of National Natural Science Foundation of China (project 62350610272) and the Department of Science and Technology of Shandong Province (Grant KY0020240040).

\bibliography{Bibliography}
\end{document}